\documentstyle[epsfig]{cupconf}

\ifoldfss
\else
  \ifnfssone
    \newmathalphabet{\mathit}
      \addtoversion{normal}{\mathit}{cmr}{m}{it}
      \addtoversion{bold}{\mathit}{cmr}{bx}{it}
    \newmathalphabet{\mathcal}
      \addtoversion{normal}{\mathcal}{cmsy}{m}{n}
    \else
    \ifnfsstwo
    \fi
  \fi
\fi

\newbox\grsign \setbox\grsign=\hbox{$>$} \newdimen\grdimen \grdimen=\ht\grsign
\newbox\simlessbox \newbox\simgreatbox \newbox\simpropbox
\setbox\simgreatbox=\hbox{\raise.5ex\hbox{$>$}\llap
     {\lower.5ex\hbox{$\sim$}}}\ht1=\grdimen\dp1=0pt
\setbox\simlessbox=\hbox{\raise.5ex\hbox{$<$}\llap
     {\lower.5ex\hbox{$\sim$}}}\ht2=\grdimen\dp2=0pt
\setbox\simpropbox=\hbox{\raise.5ex\hbox{$\propto$}\llap
     {\lower.5ex\hbox{$\sim$}}}\ht2=\grdimen\dp2=0pt
\def\simgreat{\mathrel{\copy\simgreatbox}}
\def\simless{\mathrel{\copy\simlessbox}}

\newcommand{\bc}{\begin{center}}
\newcommand{\ec}{\end{center}}
\newcommand{\be}{\begin{equation}}
\newcommand{\ee}{\end{equation}}
\newcommand{\beq}{\begin{eqnarray}}
\newcommand{\eeq}{\end{eqnarray}}
\newcommand{\bez}{\begin{eqnarray*}}
\newcommand{\eez}{\end{eqnarray*}}
\newcommand{\asap}{{\it Astronomy and Astrophysics\ \/}} 
\newcommand{\asaps}{{\it Astronomy and Astrophysics Suppl.\ \/}} 
\newcommand{\mnras}{{\it Monthly Not. Roy. Astron. Soc.\ \/}} 
\newcommand{\apj}{{\it The Astrophysical Journal\ \/}} 
\newcommand{\apjs}{{\it The Astrophysical Journal Suppl.\ \/}}

\def\rin{r_{in}}
\def\rc{r_{c}}

\def\lsintr{l_s^{\rm intr}}

\def\taut{\tau_T}
\def\taup{\tau_p}
\def\Tbb{T_{\rm bb}}
\def\Te{T_e}

\def\hexnumber#1{\ifcase#1 0\or1\or2\or3\or4\or5\or6\or7\or8\or9\or
 A\or B\or C\or D\or E\or F\fi }

\makeatletter
\ifx\CUP@mtlplain@loaded\undefined
\else
\fi
\makeatother
 \makeatletter
 \ifx\CUP@mtlplain@loaded\undefined
   \font\tenbmi=cmmib10 at 10pt
   \font\sevenbmi=cmmib10 at 7pt
   \font\fivebmi=cmmib10 at 5pt

   \newfam\bmifam
   \textfont\bmifam=\tenbmi
   \scriptfont\bmifam=\sevenbmi
   \scriptscriptfont\bmifam=\fivebmi
   
 \fi
 \makeatother
\ifnfsstwo

\fi
\ifnfssone

\fi
\ifoldfss

\fi

\mathchardef\varLambda="0103

\makeatletter
\ifx\CUP@mtlplain@loaded\undefined
\else
\fi
\makeatother
\makeatletter
\ifx\CUP@mtlplain@loaded\undefined
  \font\tenbms=cmbsy10
  \font\sevenbms=cmbsy10 at 7pt
  \font\fivebms=cmbsy10 at 5pt
  \newfam\bmsfam
  \textfont\bmsfam=\tenbms
  \scriptfont\bmsfam=\sevenbms
  \scriptscriptfont\bmsfam=\fivebms

  \edef\bsy@{\hexnumber\bmsfam}
  \mathchardef\bnabla="0\bsy@72
\fi
\makeatother
\def\etal{\mbox{\it et al.}}

\title[Accretion disc - corona models] 
{Accretion disc-corona models and X/$\gamma$-ray spectra 
of accreting black holes} 

\author[Juri Poutanen]%
{J\ls U\ls R\ls I\ns P\ls O\ls U\ls T\ls A\ls N\ls E\ls N}  
\affiliation{Stockholm  Observatory, SE - 133 36 \ Saltsj\"obaden, Sweden} 

\begin{document}
\ifnfssone
\else
  \ifnfsstwo
  \else
    \ifoldfss
      \let\mathcal\cal
      \let\mathrm\rm
      \let\mathsf\sf
    \fi
  \fi
\fi

\maketitle

\begin{abstract}
We discuss properties of thermal and hybrid (thermal/non-thermal) 
electron-positron plasmas in the pair and energy equilibria. 
Various accretion disc-corona models, recently proposed to explain properties 
of galactic  as well as extragalactic accreting black holes, 
are confronted with the observed broad-band X-ray and $\gamma$-ray spectra. 
\end{abstract}

\firstsection 

\section{Introduction} 

It was realized quite early that broad-band X/$\gamma$-ray spectra of 
Galactic black holes (GBHs) can be explained in terms of successive Compton 
scatterings of soft photons (Comptonization) in a hot electron cloud. 
The Comptonizing medium was assumed to be thermal with a given 
temperature, $\Te$, and a Thomson optical depth, $\tau_T$. 
The theoretical spectra were computed by analytical (Shapiro, Lightman \& Eardley 1976;
Sunyaev \& Titarchuk 1980) and Monte-Carlo methods (Pozdnyakov, Sobol' \& Sunyaev 
1983). 

The problem with such an approach is that in any specific geometry 
arbitrary combinations of $(\tau_T, \Te)$ are not possible. 
Both GBHs and Seyfert galaxies show a hardening of the spectra at $\sim 10$ keV, which
is attributed to Compton reflection (combined effect of photo-electric absorption 
and Compton down-scattering) of hard radiation from a cold material
(White, Lightman \& Zdziarski 1988; George \& Fabian 1991). 
Hard radiation, reprocessed in the cold matter, can form a significant fraction 
of the soft seed photons for Comptonization. The energy balance of the cold
and hot phases determine their temperatures and the  shape of the emerging 
spectrum (Haardt \& Maraschi 1991, 1993; Stern \etal\/ 1995b; Poutanen \& Svensson 1996). 

The situation becomes more complicated when a notable fraction of the total 
luminosity escape at energies above $\sim 500$ keV. Then hard photons 
can produce $e^{\pm}$ pairs which will be added to the background plasma. 
Electrons (and pairs) Comptonize soft photons up to
$\gamma$-rays and produce even more pairs. Thus, the radiation field, in this case, 
has an influence on the optical depth of the plasmas, which in its turn produces 
this radiation. This makes the problem very non-linear. 

Another complication appears when the energy distribution of particles starts to 
deviate from a Maxwellian. In the so called non-thermal models, relativistic
electrons are injected to the soft radiation field. The steady-state electron 
distribution should be computed self-consistently,  
balancing electron cooling (e.g., by Compton scattering and Coulomb interactions)
and acceleration, together with the photon distribution. 
The pioneering steps in solving this problem were done by Stern (1985, 1988) using 
Monte-Carlo techniques and by \cite{f86,lz87,coppi92} using the method of kinetic equations
(see Stern \etal\/ 1995a; Pilla \& Shaham 1997; Nayakshin \& Melia 1998, for recent
developments). 
Non-thermal model have been used extensively in the end of 1980s and beginning of 1990s 
for explaining the X-ray spectra of active galactic nuclei (see, e.g., 
Zdziarski \etal\/ 1990), while recently pure thermal model were preferred, since
the data show spectral cutoffs at $\sim$ 100 keV in both GBHs and Seyferts 
(Grebenev \etal\/ 1993, 1997; Johnson \etal\/ 1997). 
However, power-law like spectra extending without a cutoff up to at least 
$\sim 600$ keV, observed in some GBHs in their soft state 
(Grove \etal\/ 1997a,b), give new strength to the undeservedly forgotten non-thermal
models. 

Spectral fitting with multi-component models following simultaneously
energy balance and electron-positron pair balance, give stronger constraints 
on the physical condition in the X/$\gamma$-ray source, its size and 
geometry, presence of $e^{\pm}$ pairs, and give a possibility 
to discriminate between various accretion disc models. 
In this review, we first describe thermal 
as well as non-thermal pair models that have been used 
recently for spectral fitting of GBHs and Seyferts. 
We discuss spectral properties of 
$e^{\pm}$ plasmas in energy and pair equilibria for various 
geometries of the accretion flow. 
Separately for GBHs and Seyferts, we briefly review X/$\gamma$-ray observations.  
Then, we consider physical processes
responsible for spectral formation and confront  
phenomenological models of the accretion discs with data. 
We restrict our analysis to ``radiative'' models where radiative 
processes  and radiative transfer in realistic geometries 
are considered in details while heating and acceleration mechanisms 
are not specified. 

\section{Spectral models} 

\subsection{Thermal (pair) plasmas} 

Since spectra of both Seyferts and GBHs in their hard states 
cutoff sharply at $\sim$ 100 keV (Grebenev \etal\/ 1993, 1997; 
Zdziarski \etal\/ 1996a,b, 1997; Grove \etal\/ 1997a,b), thermal 
models are in some preference. 

\subsubsection{General properties} 

First, we consider properties of an electron (-positron) plasma  cloud 
in energy and pair equilibria 
without  assuming any specific geometry of the accretion flow. 
There are four parameters that describe the 
properties of hot thermal plasmas: 
(i) $l_h \equiv L_h\sigma_T/(m_e c^3 \rc)$, the hard compactness which 
is the dimensionless cloud heating rate; 
(ii) the soft photon compactness, $l_s$, which represents
the cold disc luminosity that enters the hot cloud (corona);
(iii)  $\tau_p$, the proton (Thomson) optical depth of the cloud
(i.e.., the optical depth due to the background electrons); and 
(iv) the characteristic temperature of the soft photons, $\Tbb$.
Here $\rc$ is the cloud size, $\sigma_T$ is the Thomson scattering cross-section. 
In this simplified description, energy and pair balance  equations have been 
solved using analytical and numerical methods 
(Zdziarski 1985; Ghisellini \& Haardt 1994; Pietrini \& Krolik 1995; 
Coppi 1992; Stern \etal\/ 1995a). 

\begin{figure}
\centerline{\epsfig{file=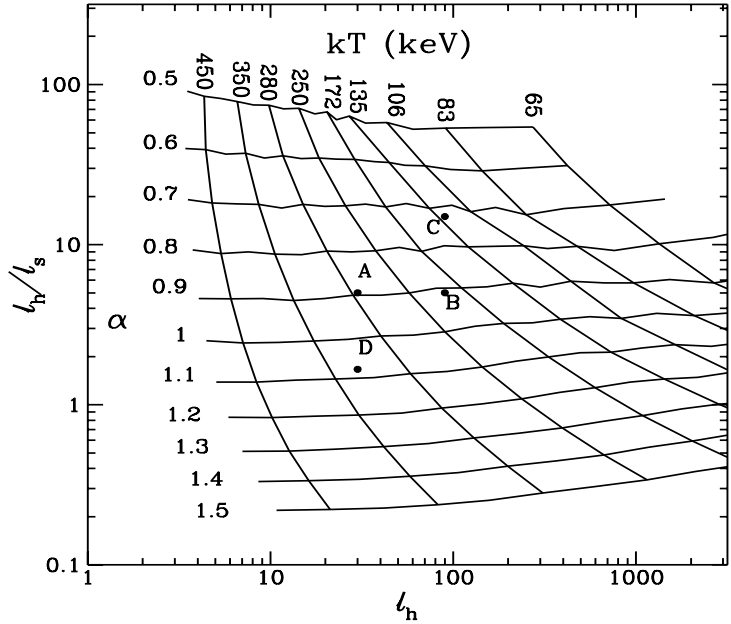,width=4.in,height=3.5in}}
\vspace{10pt}
\caption{Mapping of $(l_h,l_h/l_s)$ to $(\alpha,\Te)$
(from Ghisellini \& Haardt 1994).
The compactnesses are defined as follows:
$l_{h,s} \equiv L_{h,s}\sigma_T/(m_e c^3 \rc)$, where $\rc$ is the size
of the hot cloud, $L_h$ is the heating rate of the hot cloud, and $L_s$
is the luminosity of the seed soft photons that cool the plasma.
This relation holds for $e^{\pm}$ pairs dominated plasmas.
For electron-proton dominated plasmas, $(\alpha,\Te)$ depend
on $(\tau_T,l_h/l_s)$. One sees that ratio $l_h/l_s$ almost uniquely defines
the spectral energy index, $\alpha$.
}\label{fig:alpte}
\end{figure}

For sufficiently high compactnesses, the total optical depth can be significantly 
larger than $\taup$. In that case, increase in the cloud heating 
rate results in the corresponding increase of the cloud total optical depth, $\taut$, 
due to $e^{\pm}$ pairs produced, and in decrease of the cloud temperature, $\Te$. 
The ratio $l_h/l_s$ (the amplification factor) has a one-to-one correspondence with 
the Kompaneets $y$-parameter (e.g., Rybicki \& Lightman 1979), 
which can be related to the spectral index of the emitted 
spectrum (see Fig.~\ref{fig:alpte}). 
The spectral index stays approximately constant 
for a constant Kompaneets $y$-parameter ($y=4\Theta\tau_T$, for parameters
of interest, here $\Theta\equiv k\Te/m_ec^2$). 

Pietrini \& Krolik (1995) proposed a
very simple analytical formula that relates the observed X-ray 
spectral energy index to the amplification factor: 
\be \label{eq:alplslh}
\alpha \approx 1.6 \left( \frac{l_h}{l_s} \right)^{-1/4} .
\ee
Though the exact coefficient of proportionality  depends
on the electron temperature and the energy of seed photons,
this dependence is rather weak. 
When $l_h/l_s$ increases (the source becomes more ``photon starved''),
the observed spectrum becomes harder.

\subsubsection{Geometry} 

\label{sec:geo}

We consider here various geometrical arrangements of the hot plasma cloud and 
the source of soft photons (see Haardt 1997 for a recent review). 
The observed spectra of GBHs and Seyferts correspond to $y\approx 1$. This fact 
does not have a direct explanation in the accretion disc framework. 
If soft seed photons for Comptonization are produced by reprocessing 
hard X/$\gamma$-ray radiation, then the geometry will define the amplitude 
of feedback effect and the spectral slope (Liang 1979). 
What would be the most probable geometry? 

{\em Sandwich.} The simplest solution is to assume that a hot corona covers 
most of the cold disc (a sandwich, or a slab-corona model). The radiative 
transfer in such a geometry was considered by Haardt \& Maraschi (1991, 1993) 
who showed that 
in the extreme case, when all the energy is dissipated in the corona, the emitted
spectra resemble those observed in Seyfert galaxies. Dissipation of energy in the cold 
disc (with subsequent additional production of soft photons) would produce too 
steep spectra in disagreement with observations. Even harder spectra observed 
in GBHs cannot be reconciled with the slab-corona model 
(Dove \etal\/ 1997; Gierli\'nski \etal\/ 1997a; Poutanen, Krolik \& Ryde 1997), and 
alternative models with more photon starved conditions and smaller feedback of 
soft photons are sought. 

{\em Magnetic flares.} A patchy corona  (Galeev, Rosner \& Vaiana 1979; Haardt, Maraschi 
\& Ghisellini 1994), where the cold disk is not covered completely by 
hot material, has certainly a smaller feedback, and the resulting spectra are harder. 
A patchy corona can be described by  a number of active regions above 
the cold accretion disc. Spectral properties of an active region 
in the energy and pair balance have been computed recently by Stern \etal\/ (1995b)
(see also reviews by Svensson 1996a,b for more details). 
Both patchy and slab-corona models  predict an {\em anisotropy break} (i.e.
a break in the power-law spectrum due to the anisotropy of the seed photons) 
that should appear at the energy corresponding to 
the second  scattering order. 

{\em Cloudlets.} Another possible solution of the photon starvation problem is to 
assume that the cold disc within the hot corona is disrupted into cold dense 
optically thick clouds (Lightman 1974; Celotti, Fabian \& Rees 1992; 
Collin-Souffrin \etal\/ 1996; Kuncic, Celotti \& Rees 1997) that 
are able to reprocess hard X/$\gamma$-ray radiation and produce soft 
seed photons for Comptonization. 
If the height-to-radius ratio of the hot cloud is small, we can approximate this 
geometry by a plane-parallel slab. We assume further that the cold
material is concentrated in the central plane of the hot slab and has
a covering factor $f_c$. Compton reflection comes from these cold clouds 
(cloudlets) as well as from the outer cooler disc. 
The seed soft photon radiation is much more isotropic and the emerging 
high energy spectrum does not have an anisotropy break. 
The covering factor defines the amplitude of the feedback effect.
The total soft seed luminosity (with corresponding compactness, $l_s$) 
is the sum of the reprocessed luminosity and the luminosity intrinsically
dissipated  in the cold disc (with corresponding compactness, $\lsintr$).
For a slab geometry,  the heating rate, $L_{h}$, of a cubic volume of size $h$ 
determines the hard compactnesses $l_h \equiv L_{h}\sigma_T/(m_e c^3 h)$
(where $h$ is the half-height of the slab). Other compactnesses are defined 
is a similar way. 

\begin{figure}
\centerline{\epsfig{file=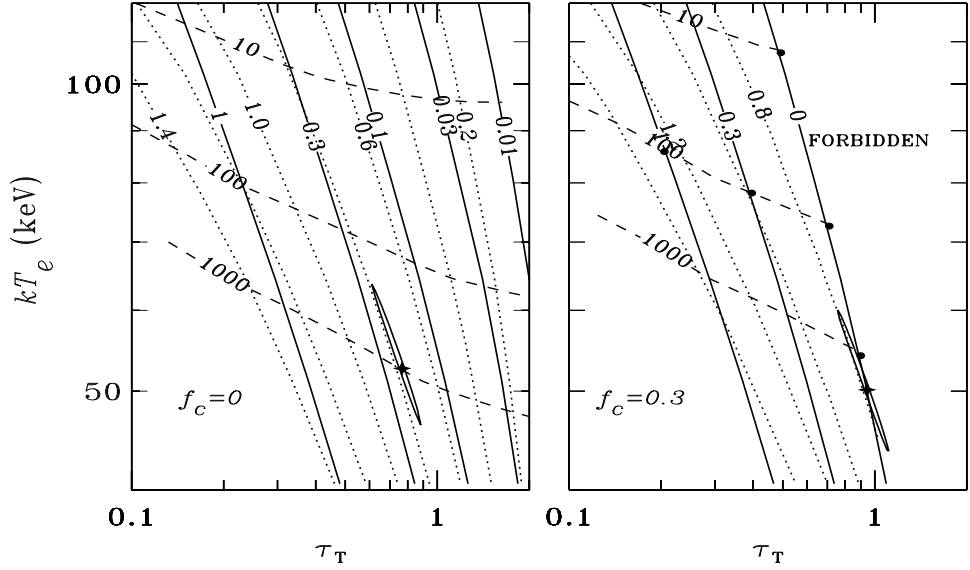,width=5.2in,height=3.in}}
\vspace{10pt}
\caption{Relation between total optical depth, $\taut$, of the half-slab 
and electron temperature, $k\Te$, for the cloudlets model.  
Here cold matter is assumed to be concentrated in the central 
plane of the hot slab. Temperature of cloudlets is fixed at 
$k\Tbb=0.25$ keV. The covering fraction of the cold material
is taken to be $f_c=0$ (left panel) and $f_c=0.3$ (right panel). 
For $f_c=0$, all seed photons are external. 
Solid curves represent solutions for constant $\lsintr/l_h$. 
These relations are the same for pure pair or pure electron-proton plasmas. 
The dashed curves are the solutions for a constant $l_h$ (assuming 
pair dominated plasmas, i.e., $\taup \ll \taut$, and 
thermal electron distribution). Dotted curves 
give solutions for a constant intrinsic (without Compton reflection) 
spectral index $\alpha$ in 2--18 keV range. 
$\lsintr$ can mean here both intrinsic 
internally (not reprocessed) and externally 
produced soft photon luminosity. 
Region to the right of $\lsintr/l_h=0$ curve is forbidden (the energy 
balance cannot be reached). 
Stars represent best fits to the simultaneous {\em Ginga} and OSSE 
data of GX 339-4 in September 1991 (Zdziarski \etal\/ 1998) and 
elongated ellipsoids are the contours plots at 90 per cent confidence level
for two interesting parameters ($\Delta\chi^2=4.61$). 
For GX 339-4, external or internally produced soft luminosity (not reprocessed)
entering hot slab can be $\sim$ 23\% of the heating rate $L_h$, if $f_c=0$. 
If $f_c=0.3$, the only solution describing data of GX 339-4 
is possible when there are {\em no} 
internally generated (except reprocessed) or external soft photons. 
Cooling is provided by reprocessed radiation only. 
Spectra for the solutions marked by filled circles are shown in 
Figure~\protect\ref{fig:slabsp}.
}\label{fig:sl025}
\end{figure}

\begin{figure}
\centerline{\epsfig{file=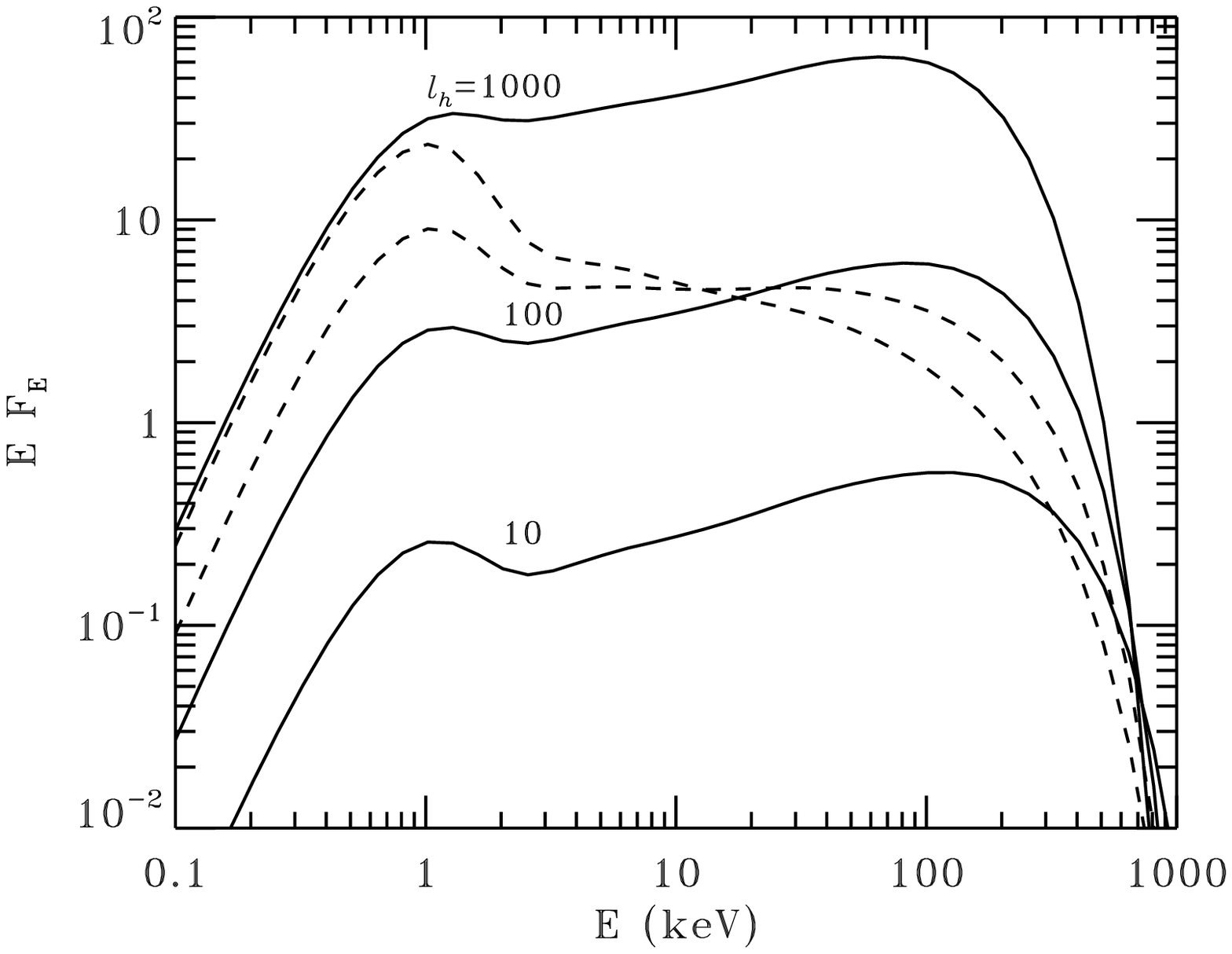,width=4.5in,height=3.5in}}
\vspace{10pt}
\caption{Spectra from the hot slab - cold clouds model for 
solutions represented by filled circles in Figure~\protect\ref{fig:sl025}. 
Solid curves correspond to the solutions with no internal dissipation 
in the cold clouds ($\lsintr=0$) and various hard compactness $l_h$. 
Dotted curves correspond to $l_h=100$, and $\lsintr/l_h=0.3$ and 1. 
With increasing $\lsintr/l_h$ spectra become steeper (softer), but the 
electron temperature does not decrease (rather slowly increases). 
In the case of the electron-proton plasmas, $\Te$ would decrease rapidly with 
constant $\taut\approx \taup$. 
}\label{fig:slabsp}
\end{figure}

Figure~\ref{fig:sl025} shows the dependence of the electron temperature
and the optical depth on parameters of the cloudlets model, and 
Figure~\ref{fig:slabsp} gives a few selected spectra. 
Using the method of \cite{ps96}, we solve the energy and pair balance  equations
coupled with the radiative transfer accounting for Compton scattering
(exact redistribution function is employed, see, e.g., Nagirner \& Poutanen 1994),
pair production and annihilation, and Compton reflection.
We should point out that in the case of pair dominated plasmas, 
increase in the amount of soft photons does not necessarily imply 
a decrease in the plasma temperature. The optical depth decreases 
rapidly with increase of $\lsintr/l_h$, and the average energy available per
particle can even increase.
In the case of electron-proton plasmas, $\taut\approx \taup$ and 
$\Te$ decreases with increasing internal dissipation in the cold disc.

{\em ``Sombrero''.} In this model, the cold disc penetrates only 
a short way into the central coronal region (see, e.g., 
Bisnovatyi-Kogan \& Blinnikov 1977, and Poutanen \etal\/ 1997 for 
recent applications).  We can assume that  the X/$\gamma$-ray source can be
approximated by a homogeneous spherical cloud of radius $\rc$ 
situated around a black hole 
(probably, a torus geometry for a hot cloud would be more physically realistic, but 
then it would be more difficult to compute the radiative transfer). 
The inner radius, $\rin$, of the cold geometrically thin, infinite
disc is within the corona ($\rin/\rc \leq 1$).
This geometry is also similar to the geometry of the popular advection 
dominated accretion flows (see article by R.~Narayan, R.~Mahadevan, and 
E.~Quataert in this volume). 
Spectra from the sombrero models are almost identical to the spectra 
expected from the cloudlets model, with the only difference that the amount 
of Compton reflection would be a bit larger for the same configuration 
of the outer cooler disc. From the observational point of view, 
these models are almost indistinguishable. 

\subsection{Hybrid thermal/non-thermal pair plasmas} 

There are reasons to believe that in a physically realistic situation, 
the electron distribution can notably deviate from a Maxwellian. A significant
fraction of the total energy input can be injected to the system in form 
of relativistic electrons (pairs). In the so called hybrid thermal/non-thermal model,
the injection of relativistic electrons is allowed in addition to the 
direct heating of thermal electrons. 

The most important input parameters of the model are:
(i) the thermal compactness, $l_{th}$, 
which characterizes the heating rate of electrons (pairs);
(ii) the analogous non-thermal compactness, $l_{nth}$, which
characterizes the  rate of injection of relativistic electrons,
(iii) the soft photon compactness, $l_s$; 
(iv) $\Gamma_{inj}$, the power-law index of the non-thermal electron
injection spectrum,
(v) $\tau_p$, the proton (Thomson) optical depth; 
and (vi) $T_{bb}$.
Compton reflection adds a few more parameters (e.g. 
the amplitude $R$, the ionisation parameter $\xi$) and can be accounted for 
using angular dependent Green's functions (Magdziarz \& Zdziarski 1995; 
Poutanen, Nagendra \& Svensson 1996).
By $l_h=l_{th}+l_{nth}$, we denote the total hard compactness.  
For spectral fitting, we use the code of \cite{coppi92} 
(see also Coppi \etal\/ 1998) which is incorporated into the standard 
X-ray data analysis software XSPEC. 

The electron distribution is computed self-consistently
balancing electron cooling (by Compton scattering and Coulomb interactions),
heating (thermal energy source), and acceleration (non-thermal energy source).
The self-consistent electron (-positron) distribution can be characterized  by a
Maxwellian of the equilibrium temperature, $T_c$, plus a non-thermal
(generally not a power-law) tail. The spectrum of escaping radiation then
consists of the incident blackbody, the soft excess due to Comptonization
by a {\em thermal} population of electrons and a power-law like tail due
to Comptonization by a {\em non-thermal} electron (pair) population.

\begin{figure}
\centerline{\epsfig{file=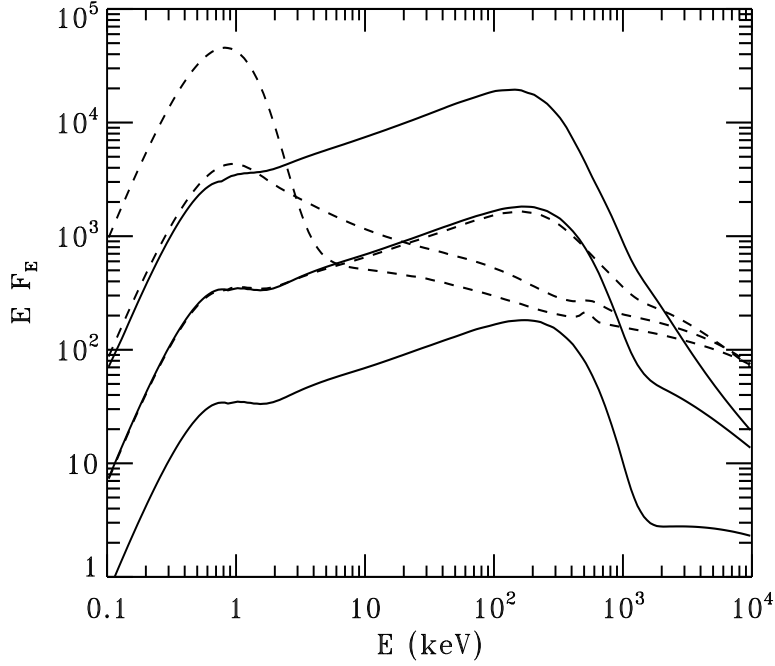,width=4.5in,height=3.5in}}
\vspace{10pt}
\caption{Spectra from the hybrid pair plasmas. 
Solid curves show dependence on hard compactness $l_h$.
Other parameters: $l_h/l_s=10$, $l_{nth}/l_h=0.1$, 
$\taup=1$, $\Gamma_{inj}=2.5$, $\Tbb=0.2$ keV. 
The resulting electron temperature and optical depth are 
$(k\Te,\taut)$=(126 keV, 1.0002), (123 keV, 1.02), and (82 keV, 1.47) for 
$l_h=1, 10, 100$, respectively ($l_h$ increases from the bottom to the top
of the figure). 
For a  higher compactness, the spectrum has a sharper cutoff at energies above 
1 MeV due to larger optical depth for photon-photon pair production. 
These spectra are similar 
to the spectra of GBHs in their hard  state (see Fig.~\ref{fig:cygx1hs}). 
Dashed curves show dependence on $l_h/l_s$. 
Here we fixed $l_h=10$, $l_{nth}/l_h=0.5$. 
The resulting electron temperature and optical depth 
$(k\Te,\taut)$ are (104 keV, 1.07), (34 keV, 1.02), and (5 keV, 1.01) for 
$l_h/l_s=10, 1, 0.1$, respectively. Increase in $l_s$ results in a more 
pronounced blackbody part of the emerging spectrum. The blackbody is modified 
by Comptonization on thermal electrons. 
}\label{fig:eqth} 
\end{figure}

As a first example, we consider how spectra from the hybrid plasmas 
change with the hard compactness when keeping the ratio
of the soft-to-hard compactness constant (this gives an almost constant $\alpha$) 
and fixing the non-thermal efficiency ($l_{nth}/l_h$) at 10 per cent 
(see solid curves in Figure~\ref{fig:eqth}). 
The electron temperature behaves exactly as in the 
pure thermal case (it decreases when compactness increases), 
since relatively small non-thermal efficiency, that we have chosen, 
does not change the energy balance significantly. 
The electron distribution is Maxwellian with a weak high energy tail. 

Next, we consider how the spectra change as a function of $l_h/l_s$, while keeping 
the other parameters constant. 
The dashed curves in Figure~\ref{fig:eqth} show the evolution of the spectrum with 
increasing soft seed photon luminosity. For large $l_h/l_s$, 
most of the spectrum is produced by Comptonization off a 
{\em thermal}  population of electrons (pairs), while the tail at 
energies above $m_ec^2$ is produced by {\em non-thermal} electrons. 
For low $l_h/l_s$, the electron temperature drops. Most of the pairs 
are in the thermal bump, but Kompaneets $y$-parameter is very small 
since the electron temperature is small. 
The resulting spectrum is produced by a 
{\em single} Compton scattering off non-thermal electrons. 
The Maxwellian part of the electron distribution produces a weak power-law 
tail to the blackbody bump. The annihilation line is quite weak 
for relatively small compactnesses, and would not be detectable by 
modern detectors. The cutoff energy at a few MeV 
is anti-correlated with the compactness.

Quantitatively, the behaviour of the electron distribution with changing 
of the amount of soft photons is easy to understand. A break between 
thermal and non-thermal parts of the electron distribution 
appears where the thermalization timescale due 
to Coulomb scattering is equal to the Compton cooling timescale 
(we neglect here thermalization by synchrotron self-absorption, see, 
e.g., Ghisellini \& Svensson 1990; Ghisellini, Haardt, \& Svensson 1998).  
Compton cooling timescale is simply $t_{Compton}=\pi r_c/(\gamma c l_s)$, 
while Coulomb thermalization  operates at 
$t_{Coulomb}\approx \gamma r_c/(\taut c \ln\Lambda)$
timescale (see, e.g., Dermer \& Liang 1989; Coppi 1992; Ghisellini, Haardt, 
\&  Fabian 1993;
here $\ln \Lambda$ is the usual Coulomb logarithm, typically $\sim$ 15, and 
$\gamma$ is the electron Lorentz factor). 
These relations define the Lorentz factor of the break 
\be 
\gamma_{break} \approx \left( \pi \ln \Lambda \frac{\taut}{l_s} \right)^{1/2} . 
\ee 
Increase in Compton cooling causes the break to shift towards lower 
energies. 

\section{Galactic black holes} 

\subsection{Hard state of Galactic black holes} 

\subsubsection{Observations and interpretation} 

\begin{figure}
\centerline{\epsfig{file=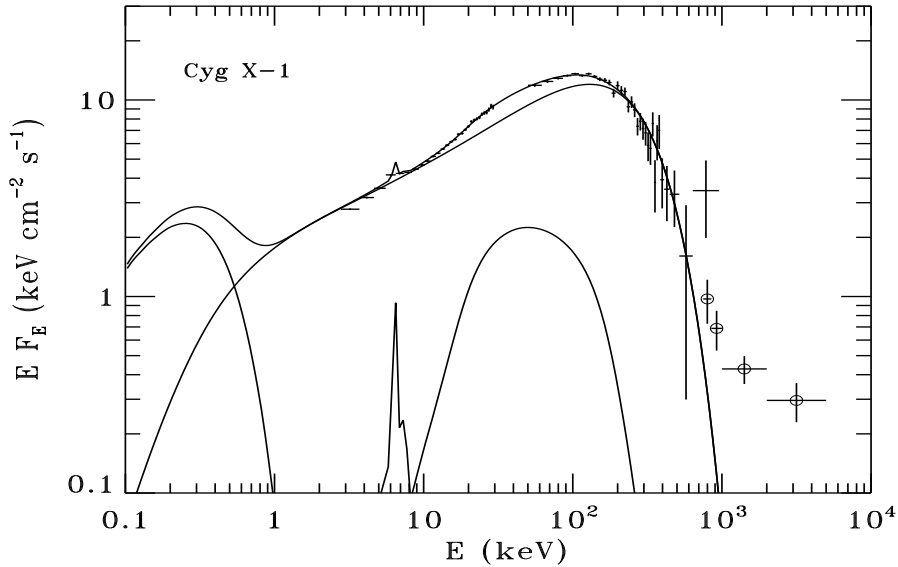,width=5.2in,height=3.in}}
\vspace{10pt}
\caption{The hard state of Cygnus X-1. Simultaneous 
{\em Ginga}, OSSE and COMPTEL data from June 1991. 
Different components correspond to the soft blackbody radiation, 
thermal Comptonized spectrum, and its Compton reflection. 
A high energy excess at $\simgreat$ 1 MeV cannot be described 
by thermal models. 
}\label{fig:cygx1hs} 
\end{figure}

Galactic black holes (GBHs) are observed in a few different spectral states 
that can be generally classified as soft and hard. 
A spectrum in the hard state is characterized by a power-law with 
the energy spectral index $\alpha\approx 0.4-0.9$ with a cutoff at energies 
$\sim$ 100 keV (Grebenev \etal\/ 1993, 1997; Tanaka \& Lewin 1995; 
Phlips \etal\/ 1996; Zdziarski \etal\/ 1996a, 1997; Grove \etal\/ 1997a,b). 
The presence of an iron line at $\sim 6.4$ keV and an iron edge at $\sim 7$ 
keV, together with a spectral hardening around 10 keV, was interpreted as 
a signature of Compton reflection of the intrinsic spectrum from 
relatively cold matter (Done \etal\/ 1992; Ebisawa \etal\/ 1996b; 
Gierli\'nski \etal\/ 1997a). 
The amount of Compton reflection
$R\equiv\Omega/2\pi\approx 0.3-0.5$  ($R=1$ corresponds to an 
isotropic X/$\gamma$-ray source atop an infinite cold slab). 
An excess at energies $\simless 1$ keV is interpreted as radiation 
from  the accretion disc (Ba{\l}uci\'nska \& Hasinger 1991; 
Ba{\l}uci\'nska-Church \etal\/ 1995) with a characteristic 
temperature, $\Tbb$, of order 0.1 -- 0.3 keV (usually quite 
poorly determined, due to strong interstellar absorption in that 
spectral range). 
Observations by BATSE and COMPTEL revealed also a presence
of a high energy excess at $\simgreat$ 500 keV in some GBHs 
(Cyg X-1: McConnell \etal\/ 1994; Ling \etal\/ 1997; GRO J0422+32: 
van Dijk \etal\/ 1995). The monochromatic luminosity, $E L_{E},$ 
peaks at about $\sim 100$ keV. A characteristic spectrum of the hard state 
GBH is shown in Figure~\ref{fig:cygx1hs}. 

GBHs show variability in X/$\gamma$-rays 
on all possible time scales, from milliseconds 
to years. The size of the emitting region cannot be much larger than 
the minimum variability time scale $\times c \approx$ 
1 ms $\times c = 300$ km.  In the context of accretion onto a black hole
it is 10 gravitational radii, $R_g\equiv GM/c^2$, (for a 10 $M_{\odot}$ black hole), 
i.e., the inner part 
of the accretion disc where most of the energy is liberated.
Since, in the hard state, GBHs radiate a big fraction of the energy in 
the hard X-rays/soft $\gamma$-rays (see Figs.~\ref{fig:cygx1hs}, \ref{fig:gbhspst}), 
the region responsible for the production of this radiation lies 
within $\sim 20 - 50 R_g$. 
The most efficient cooling mechanism responsible for formation of the 
spectra is probably thermal Comptonization of soft photons 
in the $\sim 50-100$ keV electron cloud and $\tau_T \sim 1$ 
(Shapiro \etal\/ 1976; 
Zdziarski \etal\/ 1996a, 1997; Gierli\'nski \etal\/ 1997a). High sensitivity 
OSSE observations in the 50-500 keV range allow to 
determine the electron temperature to within 10 per cent. 
High energy ($\simgreat$ 300 keV) 
excesses can be explained only if one introduces an additional 
spectral component. This component can be produced either in a spatially 
separated, much hotter region by thermal Comptonization 
(Liang \& Dermer 1988; 
Liang 1991; Ling \etal\/ 1997), or in the same region by a 
non-thermal tail of the electron distribution (Li, Kusunose \& Liang 1996a,b;
Poutanen \& Coppi 1998). In the former (thermal) case, 
the very hot $\sim 400$ keV plasma cloud has to be kept far from the 
sources of soft photons to avoid cooling, and it is not clear whether 
one can physically separate it from the rest of the accretion disc. 
Having a very hot cloud close to the $\sim 100$ keV inner disc, could also 
be a problem since radiative conduction would smooth out large
temperature gradients. On the other hand, non-thermal tails can be 
created by magnetic processes which likely operate 
in the accretion disc environment 
(see Dermer, Miller \& Li 1996; Li \etal\/ 1996a; Li \& Miller 1997 and 
references therein). 

\subsubsection{Geometry} 

Observations of the  Compton reflection feature, together with a 
fluorescent iron line, suggest the presence of a rather cold weakly ionized 
material in the vicinity of the X/$\gamma$-ray source. From the 
amplitude of  Compton reflection one can derive that the cold matter 
occupies $\Omega/2\pi\sim 1/3$  solid angle as viewed from the hard photon source 
(Ebisawa \etal\/ 1996b; Gierli\'nski \etal\/ 1997a; \.Zycki, Done \& Smith 1997a). 
Further constraints on the geometry, e.g., the covering fraction of the hot cloud
as viewed from the soft photon source, can be derived from 
the observed spectral slopes if  the {\em observed} soft luminosity
can be reliably estimated (Poutanen \etal\/ 1997).
There are no evidence that the cold material extends close to the 
black hole. (The observed iron lines are quite narrow, and the iron edges  
are quite sharp implying weakness of gravitational and Doppler effects.) 

{\em Sandwich.} 
As we pointed out in Section~\ref{sec:geo}, an accurate treatment of the 
radiative transfer and Compton scattering rule out slab-corona 
(sandwich) models for sources with hard spectra 
(see Poutanen \etal\/ 1997; Dove \etal\/ 1997 where the case of Cyg X-1 is 
considered,  and Zdziarski \etal\/ 1998 for the interpretation of GX 339-4 data). 
Even if all the energy is dissipated in the corona, the predicted spectra 
are too steep and cannot be reconciled with observations. 
Energy dissipation in the cold disc and the corresponding increase of the 
amount of seed soft photons worsen the discrepancy. 
The anisotropy break expected at a few keV in the sandwich model was never observed. 

{\em Flares.} 
Magnetic flares (active regions) on the surface of the cold accretion disc 
also predict anisotropy break in disagreement with observations. 
Active regions atop the cold disc give the right amount of Compton reflection 
(since at $\tau_T\sim 1$, a big fraction of it is scattered away), but 
produce too steep spectra as in the case of the sandwich model.
Detached active regions (Svensson 1996a,b) produce spectra with the right spectral 
slopes while predicting too much reflection ($R\sim 1$). 

{\em Cloudlets.} 
Zdziarski \etal\/ (1998) found this model giving the best description
of {\em Ginga} and OSSE data of GX 339-4. In this case, most of the Compton reflection 
occurs in an outer cold disc. 
The covering factor of cold clouds within the hot inner disc 
cannot be larger than $f_c\sim 0.3$ 
due to the energy balance requirements (see Fig.~\ref{fig:sl025}). 

{\em Sombrero.} 
This geometry is  consistent 
with the amount of Compton reflection observed in GBHs in their hard state 
(Dove \etal\/ 1997; Gierli\'nski \etal\/ 1997a; Poutanen \etal\/ 1997).
In the case of Cyg X-1, a solution with $\rin/\rc=1$ 
is energetically possible, but the intrinsic soft luminosity should be 
rather large in order to produce enough soft photons for Comptonization. 
The energy balance constrains the inner radius of the cold disc to be 
larger than $\sim 0.7 \rc$. 
Probably, $\rin/\rc=0.8-0.9$ would satisfy all the 
observational requirements (see Poutanen \etal\/ 1997). 
Similarly, in case of GX 339-4 (which has a steeper spectrum in the hard 
state than Cyg X-1), $\rin/\rc \geq 0.7$ required. 

Concluding, models with the central hot cloud surrounded by a cold disc 
give the best description of the data. 

\subsubsection{Spectral variability and $e^{\pm}$ pairs} 

\begin{figure}
\centerline{\epsfig{file=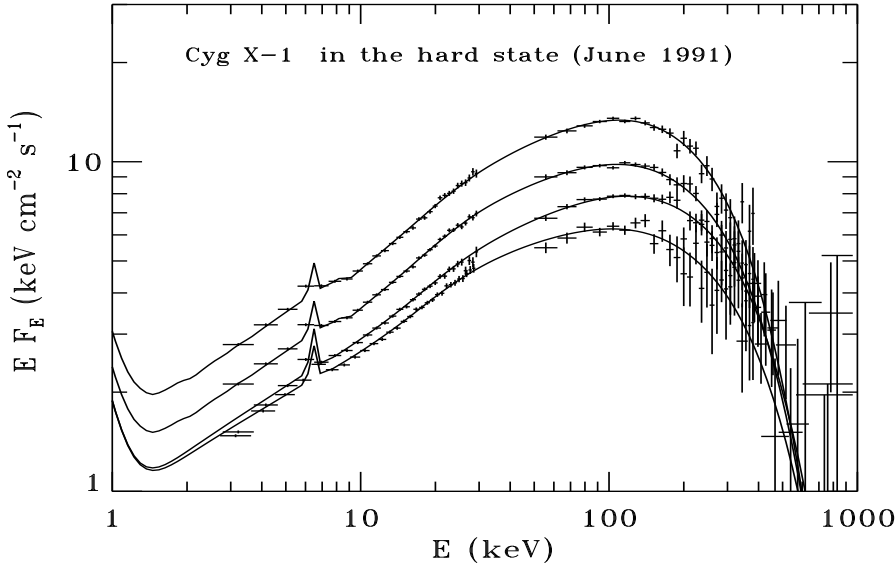,width=5.2in,height=3.in}}
\vspace{10pt}
\caption{Spectral variation of Cyg X-1 on the time scales of hours 
in the hard state as observed by {\em Ginga} and OSSE in June 1991
(data are adapted from Gierli\'nski \etal\/ 1997a). 
}\label{fig:cygx1hs4} 
\end{figure}

As it was already mentioned, Galactic black holes show variability on different 
times scale (see recent review by Van der Klis 1995).  Here we just consider
spectral variations on the time scales of hours. It was shown 
by \cite{gier97a} that the spectral shape of Cyg X-1 in the {\em Ginga} spectral range 
does not vary much when luminosity changes within a factor of two 
(Fig.~\ref{fig:cygx1hs4}), while there is evidence that the cutoff energy increases
when luminosity drops (best fit curves cross each other at $\sim$ 500 keV). 
Such a behaviour implies almost a constant ratio $l_s/l_h$ (see Figs.~\ref{fig:alpte} 
and \ref{fig:sl025}) 
and a  constant Kompaneets $y$-parameter. We can conclude that the 
transition radius between the hot and the cold discs 
does not change much (otherwise, the ratio of the intrinsic energy 
dissipation in the cold disc to the heating rate of the hot cloud would 
change, causing spectral slope changes). 
Alternatively, the transition radius is sufficiently large that 
most of the seed soft photons are provided by reprocessing hard radiation 
from the central cloud. 

For $e^{\pm}$ pair dominated plasmas, variation in the heating rate 
would cause variations in the optical depth by pair production and   
corresponding changes of the cloud temperature. On the other hand, the 
same behaviour is expected when no pairs are present. Now, variations in the 
accretion rate cause variation in optical depth and escaping luminosity, and 
the temperature adjusts to satisfy the energy balance. 

It is probably not possible to determine the pair content directly from observations, 
since spectra from the {\em thermal}  $e^{\pm}$ plasma are 
indistinguishable from spectra of the electron-proton plasma (annihilation line is 
much too weak to be observed,
see Macio{\l}ek-Nied\'zwiecki, Zdziarski \& Coppi 1995; Stern \etal\/ 1995b). 
It is in principle possible to determine the compactness parameter from 
observations. For example, in the case of Cyg X-1, $l_h\sim 20$ assuming 
a unitary (i.e. not broken into many pieces) source, while $l_h \sim 400$ is 
required to make the source pair dominated (Poutanen \etal\/ 1997). 
If the energy dissipation is extremely inhomogeneous (which would be the case if 
magnetic reconnection is responsible for the energy dissipation), 
then  at a given moment most of the luminosity is produced by a smaller fraction
of the cloud and the effective compactness can be much higher. 
Presence of non-thermal particles in the source would increase pair production 
and explain the high energy excess at $\sim 1$ MeV at the same time. However, 
the quality of the data is not sufficient to make any definite conclusions.

\newpage

\subsection{Soft state of Galactic black holes}

\subsubsection{Observations and interpretation} 

\begin{figure}
\centerline{\epsfig{file=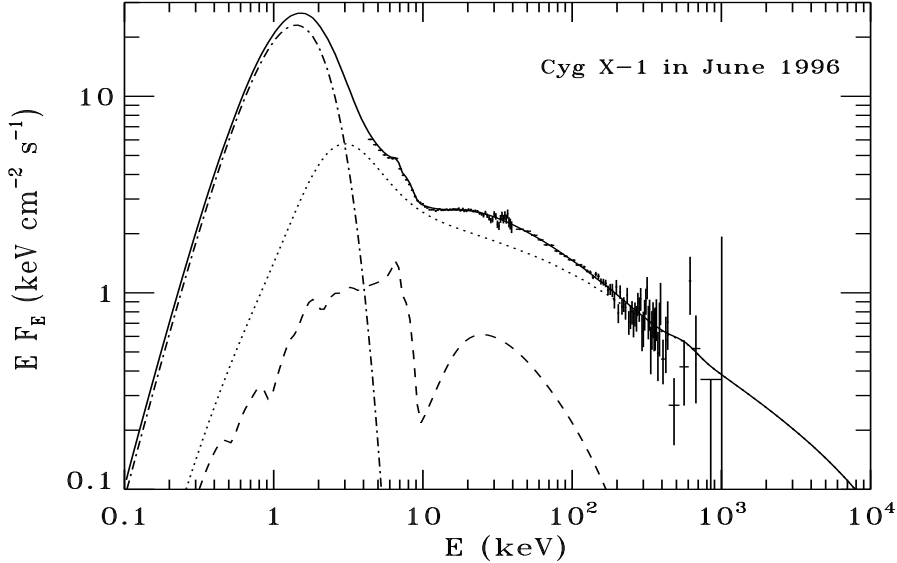,width=5.2in,height=3.in}}
\vspace{10pt}
\caption{The soft state of Cygnus X-1 observed by {\em RXTE} and OSSE in 
June 1996 and the best fit
hybrid thermal/non-thermal pair model (corrected for interstellar 
absorption). 
The solid curve represents the total 
spectrum, the dashed curve gives the Compton reflection spectrum, and the dotted curve
represents the Comptonized continuum. The disc blackbody is shown by the dot-dashed curve. 
The parameters of the fit are: $l_{s}=20$ (frozen), $l_h/l_s=0.3$, $l_{nth}/l_h=0.95$,
$\taup=0.3$, $k\Tbb=0.36$ keV, $\Gamma_{inj}=3.0$, $R=0.4$, $\xi=3.7\cdot 10^3$, 
giving a $\chi^2$/dof=168/167. 
The temperature of the Maxwellian part of the electron distribution 
is $k\Te=30$ keV, and the total Thomson optical depth (including pairs)
$\taut=0.32$. 
The iron edge and the iron line appear to be smeared (requiring to account for 
rotation of the relativistic accretion disc, see \.Zycki \etal\/ 1997a,b; 
Gierli\'nski \etal\/ 1997b) and the reflector to be ionised. 
}\label{fig:sshybr}
\end{figure}

Unlike in the hard state, most of the luminosity in the soft state is carried
by a blackbody like component with a characteristic temperature, 
$k\Tbb\approx 0.5-1$ keV. Until recently, there were not so many broad-band 
data with high spectral resolution and high signal-to-noise ratio 
(see, e.g., Tanaka \& Lewin 1995; Grebenev \etal\/ 1993, 1997) that a detailed
spectral analysis would be possible. Having {\em ASCA, RXTE}, and {\em CGRO}
in orbit at the same time changes the situation. Cygnus X-1 was observed 
by all these observatories simultaneously on May 30, 1996 and by 
{\em  RXTE} and {\em CGRO} on June 17-18, 1996, 
when it was in the soft state. 
Figure~\ref{fig:sshybr} gives an example of the soft state spectrum.  

A soft component (energies less than $\sim 5$ keV) 
cannot be represented neither by 
a multicolor disc spectrum, nor a modified blackbody. It is clear 
that at least two component are required to fit it (e.g., blackbody and 
a power-law, or two black bodies, Cui \etal\/ 1997a,b; 
Gierli\'nski \etal\/ 1997b). 
One can imagine that the soft blackbody comes from the accretion disc, but
the nature of the additional component peaking at $\sim 3$ keV is not 
so clear. \cite{gier97b} interpreted it as due to thermal Comptonization 
of a disc blackbody in a plasma with $k\Te\approx 5$ keV and $\taut\approx 3$. 

GBH spectra in the hard X-rays/soft $\gamma$-rays can be well represented by 
a power-law which does not have 
an observable break, at least, up to energies of order $m_ec^2$ 
(Phlips \etal\/ 1996; Grove \etal\/ 1997a,b, 1998). 
COMPTEL has detected Cyg X-1 and GRO J1655-40 
at energies up to $\sim 10$ MeV (A.~Iyudin, private communications), 
and it seems that the power-law at MeV energies is just a continuation 
of the hard X-ray power-law.  
Although signatures of Compton reflection are also observed 
in this state (e.g. Tanaka 1991), its amplitude, $R$,  
is much more difficult to determine since
it depends on the assumed distribution of ionisation and 
detailed modelling of the continuum, which is rather curved 
in the spectral region around the iron edge (see Fig.~\ref{fig:sshybr}). 
For example, \cite{gier97b} give $R\approx 0.6-0.8$ for Cyg X-1 in the 
soft state observed on May 30, 1996, while \cite{cui97b} get $R\approx 0.15$
(restricting themselves to a much narrower energy range). 
Both, the iron line and the iron edge, appear to be smeared due to probably 
gravitational redshift and Doppler effect, implying that the cold disc 
extends very close to the central black hole. 

The origin of the steep power-law was interpreted 
in terms of bulk Comptonization in a converging flow 
(Ebisawa, Titarchuk, \& Chakrabarti 1996;
Titarchuk, Mastichiadis, \& Kylafis 1997). 
This model predicts a cutoff at $\simless m_ec^2$ which does not appear 
to be the case. 
The power-law can be produced by Comptonization of soft 
photons from the accretion disc by a non-thermal corona (the base of the jet?) 
which is optically thin and covers much of that disc (Mineshige, Kusunose 
\& Matsumoto 1995; Li \etal\/ 1996a,b; Li \& Miller 1997; 
Liang \& Narayan 1997; Poutanen \& Coppi 1998). In that case, the turnover is expected 
at a few MeV due to absorption by photon-photon pair production.

\subsubsection{Hybrid pair plasma model} 

The steep power-law in the $\gamma$-ray spectral region can be interpreted  
as a Comptonization (or, in fact, a single scattering) 
by {\em non-thermal} electrons ($e^{\pm}$ pairs),  
and an additional component peaking at $\sim 3$ keV as a {\em thermal} 
Comptonization by rather low temperature electrons. 
It is natural to assume that both components are produced in the 
same spatial region by electrons having a non-Maxwellian distribution. 

The soft state data are shown on
Figure~\ref{fig:sshybr} together with the model spectrum. 
Unfortunately, data in the $\gamma$-rays are not 
good enough to determine the compactnesses unambiguously (one needs very accurate
estimates of the amplitude of the annihilation line, as well as 
shape of the cutoff at $\sim 5-10$ MeV). 
It is worth mentioning once more that the whole  spectrum from soft
X-rays to $\gamma$-rays can be represented by a hybrid model where the electron
distribution is determined self-consistently by balancing heating, acceleration and
cooling.

\subsection{Spectral state transitions} 

\begin{figure}
\centerline{\epsfig{file=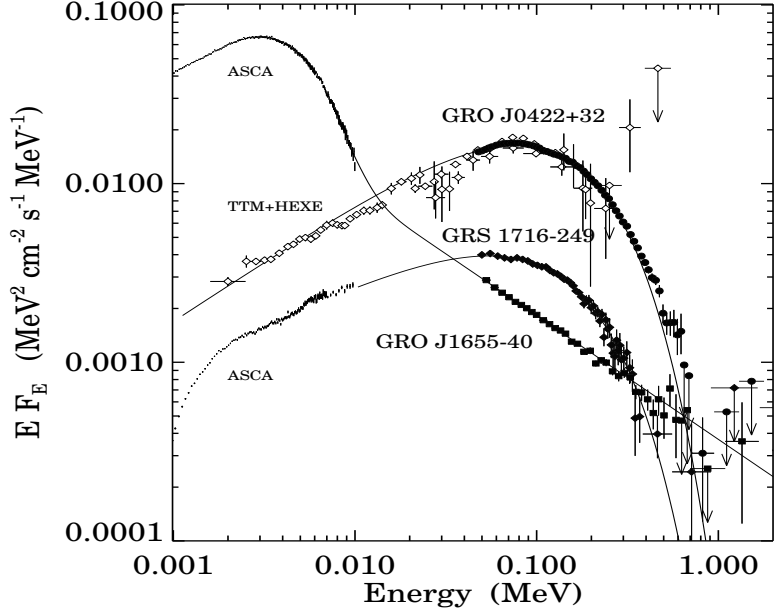,width=4.3in,height=3.5in}}
\vspace{10pt}
\caption{Different spectral classes of Galactic black holes (from 
Grove \etal\/ 1998). 
}\label{fig:gbhspst} 
\end{figure}

Some Galactic black holes have been observed to always be in one of the states 
(either in the hard, or in the soft), 
while others have shown transitions between states (see Fig.~\ref{fig:gbhspst}, 
and Sunyaev \etal\/ 1991; Grove \etal\/ 1997b; Grebenev \etal\/ 1997). 
The nature of the state transitions is not fully understood yet, and 
none of the dynamical accretion disc models can fully describe them. 
Probably, the most developed model presently available is the advection dominated
accretion disc model (see Esin \etal\/ 1997, 1998, and article by R.~Narayan \etal\/ 
in this volume). This model is able 
to explain general spectral behaviour in the X-ray range, during the transition. 
However, restricted to pure thermal plasma, it is not able to explain 
the $\gamma$-ray data. 

We restrict our consideration here to the observations and spectral 
modelling of Cyg X-1. 
During the transition from the hard to the soft state, 
the $\gamma$-ray luminosity drops, 
the spectrum becomes steeper, while the cutoff energy increases (Phlips \etal\/ 1996; 
Poutanen 1998).   
The luminosities of the different spectral components change dramatically 
in expense of each other, while the bolometric luminosity 
hardly changes (Zhang \etal\/ 1997). 

In the hard state, most of the power is deposited, through the thermal
channel, to heat the plasma of the inner accretion disc. The non-thermal 
supply is relatively small, resulting in a weak observed tail at 
MeV energies. 
The soft photon input to the system is also rather small. 
Thus the system is ``photon starved'' and produces a hard spectrum. 
Soft photons that are Comptonized to X/$\gamma$-rays by {\rm thermal} electrons 
are most probably  produced  by reprocessing of the X/$\gamma$-rays 
in the cold material. This feedback effect fixes the spectral index 
at a value that is defined by the geometry of the system. 

\begin{figure}
\centerline{\epsfig{file=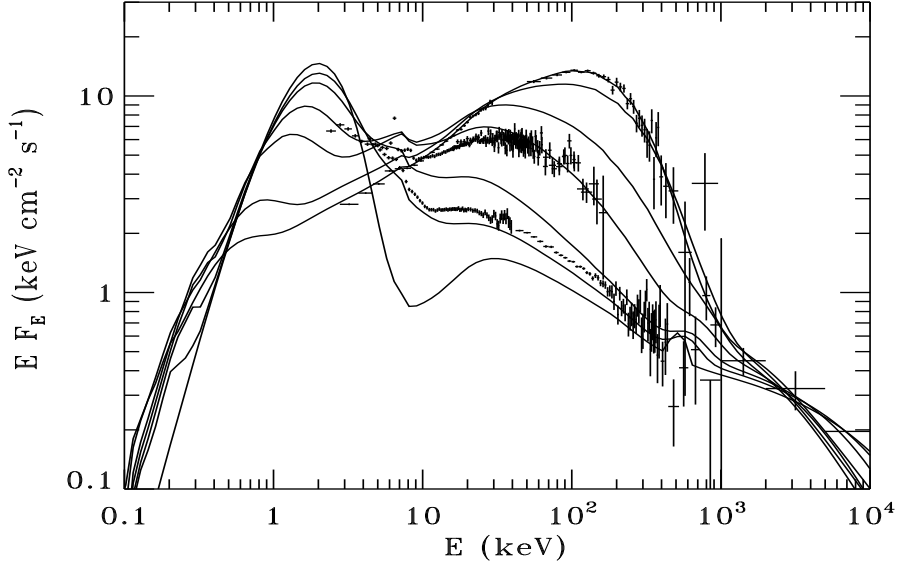,width=5.2in,height=3.in}}
\vspace{10pt}
\caption{Spectral states of Cyg X-1 and the state transition as 
predicted by the hybrid pair model. 
}\label{fig:hybrid} 
\end{figure} 

In the soft state, the soft photon luminosity (from the accretion disc) 
exceeds all other energy injection rates to the system, 
while thermal energy dissipation (electron heating in the corona/hot disc) is negligible. 
The non-thermal energy injection rate is $\sim 1/4$ of the total energy 
output of the system. This almost pure non-thermal model 
reproduces the broad-band soft state data (see Fig.~\ref{fig:sshybr} and
Gierli\'nski \etal\/ 1998). 

Poutanen \& Coppi (1998) 
showed that by using simple scaling laws for the luminosity of the cold disc, 
the thermal dissipation/heating rate in the corona, and the rate of energy 
injection from a non-thermal source, all as functions of radius of the corona, 
the hard-to-soft transition  can be explained as the result of a decrease in the
transition radius between the inner hot disc (corona) and the outer cold disc by a 
factor $\sim$ 5. 
They assumed that the sum of the soft luminosity from the disc, $L_s\propto 1/r$, and
the thermal dissipation rate in the corona, $L_{th}\propto 1-1/r$, as well as 
the non-thermal power, $L_{nth}$, remain approximately constant during transition. 
This idea is somewhat similar to the proposal by \cite{mine95} (although they 
fixed the ratio $l_{nth}/l_{th}$). 
The model gives a sequence of spectra, shown in 
Figure~\ref{fig:hybrid} together with the data of Cyg X-1. 
When transition radius decreases, the ratio $l_h/l_s$ (and the spectral index) 
does not change 
until the internally generated soft photon luminosity becomes comparable to the
reprocessed one. After that the spectrum changes dramatically since $l_h/l_s$ decreases.  
The model explains the pivoting behaviour of the spectra at $\sim 10$ keV, and predicts 
a pivoting behaviour at $\simgreat 1$ MeV. Smaller soft $\gamma$-ray luminosity
in the soft state results in a decrease of the pair production opacity, giving
a higher cutoff energy. The predicted behaviour in the MeV range 
can be compared with observations only after the launch of INTEGRAL.

\section{Seyfert galaxies} 

Broad-band spectral properties of Seyfert (Sy) galaxies are discussed in recent 
reviews by \cite{zdz97,joh97} (see also a review by G.~Madejski in this 
volume). 

\subsection{``Normal'' Seyfert 1 galaxies} 

\subsubsection{Observations}

\begin{figure}
\centerline{\epsfig{file=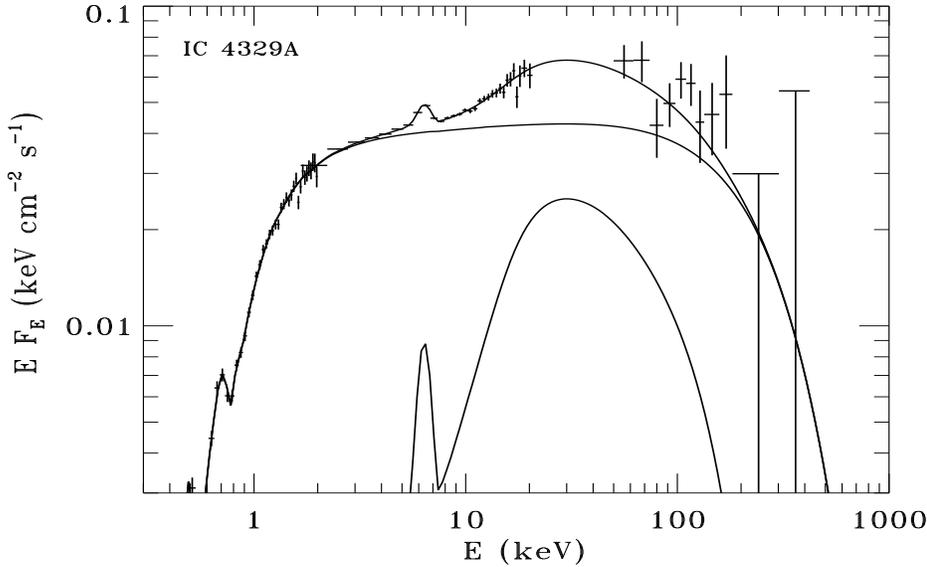,width=5.2in,height=3.in}}
\vspace{10pt}
\caption{The ROSAT/{\em Ginga}/OSSE broad-band spectrum of Seyfert 1 galaxy, IC4329A
(data from Madejski \etal\/ 1995). The dashed line is the thermal 
Comptonization spectrum from a spherical cloud of electron temperature 
$k\Te=85$ keV and radial optical depth $\tau=1.1$. The temperature of 
the soft seed photons, $k\Tbb$, is fixed at 35 eV. Calculation are made 
for the geometry of the X/$\gamma$-ray source where cold accretion disc
is situated outside of a hot central spherical cloud (see Poutanen \etal\/ 
1997; Dove \etal\/ 1997). The amount of Compton reflection is factor 
of 3 larger than expected from an infinite geometrically thin cold slab.} 
\label{fig:sy} 
\end{figure}

X-ray/gamma-ray spectra of Sy 1s  are very similar to the hard state 
of GBHs (see Fig.~\ref{fig:sy}), having a power-law 
spectral index $\alpha\approx 1$ and cutoff energies $E_c\approx 300$ keV 
(e.g., Gondek \etal\/ 1996). The spectral hardening 
at $\sim$ 10 keV is also detected in Seyferts (Nandra \& Pounds 1994; 
Weaver, Arnaud \& Mushotzky 1995). 
An important difference is that the amount 
of Compton reflection is generally found to be somewhat higher, $R\sim 0.8$
(Gondek \etal\/ 1996; Zdziarski \etal\/ 1997). Probably, the only 
exception is NGC 4151 which has almost an identical spectrum to the one 
of the GBHs GX 339-4 (Zdziarski \etal\/ 1998), and 
$R\approx 0.4$ (Zdziarski, Johnson \& Magdziarz 1996).  
Radio quiet Sy have not been detected in the $\gamma$-ray spectral range 
(Maisack \etal\/ 1995). 
Spectral similarities with GBHs  support the attribution of the X/$\gamma$ spectra
to a scale invariant process, such as Compton scattering. 
Best fits with thermal Comptonization models to the broad-band data 
give $k\Te\approx 100$ keV and $\taut\approx 1$.

\subsubsection{Geometry} 

The {\em slab-corona} model (Liang 1979; Haardt \& Maraschi 1991, 1993)  predict 
slopes of the Comptonized spectrum $\alpha\sim 1$, if all the energy is dissipated 
in the corona. Harder X-ray spectra can be achieved only if the 
coronal temperature is high enough ($\simgreat 300$ keV) to make an anisotropy
break in the observed band (Stern \etal\/ 1995b; Svensson 1996a,b). 
Let us consider as an example one of the brightest Sy 1s, IC 4329A. 
Spectral fitting to the broad-band data (Madejski \etal\/ 1995) 
gives $k\Te=80$ keV and $\taut=0.7$ (assuming $k\Tbb=35$ eV). 
However, this model can be  ruled out by the
requirement of energy balance. For a given $\taut$, the reprocessed flux from 
the cold slab would cool down the corona to 45 keV. Internal dissipation in the 
cold slab would worsen the discrepancy. Similar conclusion can be 
drawn for the averaged spectrum of Sy 1s (Zdziarski \etal\/ 1997). 

{\em Flares.} Since in Seyferts, the amplitude of Compton reflection, $R$, 
is closer to 1 than in GBHs, one could 
argue that active region (magnetic flares) model is more viable here. 
Indeed, having less soft photon returning to the active region, the spectra 
(harder than in the slab-corona case) satisfy the energy balance condition, and 
produce the needed amount of Compton reflection. 
The anisotropy  break (never seen in Seyferts) shifts to smaller energies, out
of the observed X-ray band, due to smaller (than in GBHs) characteristic  energies
of soft seed photons. 

{\em Cloudlets.} 
The hardness of the spectrum of IC 4329A, provides limits on the covering factor
of the cold clouds. The energy balance can be satisfied for $f_c\simless 1/3$ 
(similar to GX 339-4, see Fig.~\ref{fig:sl025}; note that curves corresponding
to a constant $l_h/l_s$ for $k\Tbb=35$ eV are shifted a little to the left). 
Larger $f_c$ would cool down the hot slab by the reprocessed radiation. 
For $f_c=1/3$ (no energy dissipation in clouds is allowed),  the best fit
to the data  gives $k\Te=70$ keV and the optical thickness of the half-slab 
$\taut=0.6$. This model requires Compton reflection to be produced externally. 

{\em Sombrero.} 
Models with a hot inner disc (modelled as a sphere) 
and a cold outer disc, give acceptable fits.  They are energetically possible
if $\rin/\rc \simgreat 0.7$. For $\rin/\rc=0.7$ the best fit 
gives $k\Te=75$ keV and $\taut=1.5$, and for $\rin/\rc=1$, $k\Te=85$ keV and $\taut=1.1$ 
(see Fig.~\ref{fig:sy}). Both, the sombrero model (where the cold disc is 
modelled as an infinite, geometrically thin, slab) 
and the cloudlets model predict too little Compton reflection. 
Probably, it can be provided by a material far away from the center and plane 
of the disc (molecular torus, see Krolik \etal\/ 1994; Ghisellini \etal\/ 1994). 
This can be checked by the response of the reflection component on changes in 
the amplitude of the continuum.

\subsection{Seyfert 2s}

Seyfert 2 galaxies have generally higher column densities of the absorber 
towards the nuclei, than Sy 1s do, so that their intrinsic (nuclear) 
radiation is not always directly 
observable (see, e.g. Smith, Done \& Pounds 1993; Done, Madejski \& Smith 1996;
Matt \etal\/ 1997). On the other hand, hard X-ray/soft $\gamma$-ray spectra 
of Sy 2s and Sy 1s are very similar 
(Zdziarski \etal\/ 1997; Johnson \etal\/ 1997) as predicted 
by the unification schemes of Seyfert galaxies (Antonucci 1993), and 
there are no really strong arguments that the intrinsic spectra 
of Sy 2s are somewhat different from that of Sy 1s. 

\subsection{Narrow line Seyfert 1 galaxies}

Another class of Seyfert galaxies, the ultra soft narrow-line Seyfert 1 
galaxies (NLSy 1), is characterized by a large soft X-ray excess 
(very steep spectrum in the soft X-rays) and
somewhat steeper (softer) spectra in the standard 
(1-10 keV) X-ray range  (see Turner \etal\/ 1993; Pounds, Done \& 
Osborne 1995, Brandt, Mathur \& Elvis 1997). Their hard X-ray properties 
are unknown.  
Pounds \etal\/ (1995) interpreted these objects as Seyferts in their 
soft (high) state. One would think that in that case, broad fluorescent 
iron lines should appear in these objects more often than in ``normal''
Seyferts, since the inner edge of the cold disc should be closer 
to the central black hole in order to produce large soft X-ray luminosity.
One can expect some indications of the 
correlations between the spectral index and the width of the iron line. 
We certainly need better data to make further progress in understanding 
these objects. 

\section{Final remarks}

One of the most intriguing developments during recent years, is the 
understanding that X/$\gamma$-ray  spectra of stellar mass black holes 
(in their hard states) 
are very similar to the spectra of active galactic nuclei, that are believed
to contain $10^6-10^8$ solar masses. The spectral fits to the data with 
the thermal Comptonization  models give values of the 
electron temperature within 50--100 keV and a Thomson optical depth 
(of the slab) close to unity.
These observations give more support to the scale free accretion disc models 
of the central engine and to Comptonization as the most important scale free 
radiative process. There are, however, a few questions that have to be addressed. 

What is the physical reason for $\tau_T$ to be $\sim 1$? 
If the plasma is $e^{\pm}$ pair dominated, $\tau_T\sim 1$ is a natural 
limit just because it is difficult to get compactnesses 
larger than $\sim 10^3$. Such a compactness is at least an order of magnitude 
above the estimate, given by the known X/$\gamma$-ray luminosity and sizes inferred 
from the inner edge of the outer cold disc. Significant contribution from 
non-thermal processes and inhomogeneous energy dissipation probably 
can remove this discrepancy. On the other hand, $\tau_T\sim 1$ can be 
achieved in the hot accretion discs radiating at the maximum accretion 
rate limited by advection (Zdziarski 1998).  

What is the geometry of the X/$\gamma$-ray producing region in GBHs and Seyferts? 
Magnetic flares above a cold accretion disc are still a possible solution for Seyferts. 
We argued that in the case of GBHs, a hot inner disc with a cold 
disc outside is a more plausible geometry. This can also be a {\it unifying 
geometry for both GBHs and Seyferts}. Larger amplitude of the 
Compton reflection observed in Seyferts can be due to the contribution from 
the molecular torus. On the other hand, the presence of the gravitationally 
redshifted fluorescent iron lines in the spectra of some Seyferts 
(implying inner radius of the cold disc to be at a few $GM/c^2$, see, 
e.g., Tanaka \etal\/ 1995; Fabian \etal\/; Nandra \etal\/ 1997) would be 
more difficult to explain in such a geometry. 

An important observational progress, made during recent years, was the 
broad-band X/$\gamma$-ray data of GBHs in their soft state. 
Luckily for us, the soft state
transition was observed in Cyg X-1 in the summer of 1996. 
These observations revealed that the soft state spectra cannot be explained
by thermal Comptonization. 
Hybrid thermal/non-thermal model gives an acceptable description 
of the broad-band data (from soft X-ray to MeV), 
making predictions for the spectral change in the MeV range. 
Unfortunately, we will have to wait for INTEGRAL, before we can verify these
predictions. 
This model also successfully reproduces 
spectral transitions, as a results of redistribution of the 
energy dissipation between the hot inner cloud and the cold outer disc, with a 
constant non-thermal energy injection,  
probably by the base of the jet or non-thermal corona.
Thus, the hybrid model can be a {\it unifying link between hard and soft 
states of GBHs}.
What is the physical reason for the change in the transition radius between hot and 
cold phases? We do not know the answer yet. 

In the soft state, the inner radius of the cold disc moves closer to the 
central black hole. The profile of the iron line should change notably 
and smearing of the iron edge is expected due to the Doppler effect and 
gravitational redshift. Simultaneous data with a high  spectral resolution  
and a broad spectral coverage (to determine continuum unambiguously)
are required to quantify the amplitude of these effects.  
Observationally it is a challenge, since the soft blackbody bump (in GBHs) 
dominates in the spectral region around the iron line. 

In the case of Seyferts, it is possible that 
those objects that show redshifted iron lines 
belong to the NLSy 1 class
(probably, the ``soft state'' Seyferts, see, e.g.,  Lee \etal\/ 1998 for 
the case of  MCG-6-30-15).  
Then, the correlation between the spectral index and the width of the iron line 
is expected. 
We can speculate that NLSy 1 are analogous to the soft state GBHs, but 
observationally this is not well established yet.

\begin{acknowledgments}
This research was supported by grants from the Swedish Natural Science 
Research Council and from the Anna-Greta and Holger Crafoord's Fund. 
The author thanks A.~Zdziarski, M.~Gierli\'nski, R.~Svensson, E.~Grove,
and F.~Haardt for various 
help during preparation of this review. I also would like to thank the organizers of 
the Symposium on Non-Linear Phenomena in Accretion Discs around Black Holes 
for the financial support. 
\end{acknowledgments}


\end{document}